# IMPROVEMENT OF LOADABILITY IN DISTRIBUTION SYSTEM USING GENETIC ALGORITHM

Mojtaba Nouri[1], Mahdi Bayat Mokhtari[2], Sohrab Mirsaeidi [3], Mohammad Reza Miveh[4]

[1]Department of Electrical Engineering, Saveh Branch, Islamic Azad University, Saveh, Iran,
`mojtaba.nuri@yahoo.com`

[2]Department of Electrical Engineering, Saveh Branch, Islamic Azad University, Saveh, Iran,
`mbayatm2004@yahoo.com`

[3] Young Researchers Club, Saveh Branch, Islamic Azad University, Saveh, Iran,
eng@sohrabmirsaidi.com

[4] Young Researchers Club, Saveh Branch, Islamic Azad University, Saveh, Iran,
`mivem@yahoo.com`

*Abstract* - *Generally during recent decades due to development of power systems, the methods for delivering electrical energy to consumers, and because of voltage variations is a very important problem ,the power plants follow this criteria. The good solution for improving transfer and distribution of electrical power the majority of consumers prefer to use energy near the loads .So small units that are connected to distribution system named "Decentralized Generation" or "Dispersed Generation". Deregulated in power industry and development of renewable energies are the most important factors in developing this type of electricity generation. Today DG has a key role in electrical distribution systems. For example we can refer to improving reliability indices, improvement of stability and reduction of losses in power system. One of the key problems in using DG's, is allocation of these sources in distribution networks. Load ability in distribution systems and its improvement has an effective role in the operation of power systems. However, placement of distributed generation sources in order to improve the distribution system load ability index was not considered, we show DG placement and allocation with genetic algorithm optimization method maximize load ability of power systems .This method implemented on the IEEE Standard bench marks. The results show the effectiveness of the proposed algorithm .Another benefits of DG in selected positions are also studied and compared.*

## 1-INTRODUCTION

As predicted the distributed generation will have a growing role in the future of the power systems, and also in recent years, this role has been slowly increasing [1]. "International Atomic Energy Agency" (IAEA), offers following definition for distributed generation: The units of generation that give service to the customer in the place[2]. "International Council on Large Electric Systems"(CIGRE), offers following definition for distributed generation[1]:

DOI : 10.5121/acij.2012.3301                                                                                                                                    1

1 -Not to be planned centrally

2 - Not to be transferred centrally

3- Usually is connected to the distribution network

4 - Capacity between 50 to100 MW

However, the best definition for DG is, "the source of electric energy is connected to distribution networks or directly to the consumer side". The nominal amounts of these generations varied, but usually their generation capacity range from a few KW to 10 MW. These units are in substations and in the distribution feeders, near the loads. The effects of DG on the voltage profile, line losses, short circuit current, the amount of harmonic injection, stability and reliability of the network before installation should be evaluated. DG placement and size are very important, because it's non-optimal installation increases the losses and rises the costs. Therefore to considering the above items and consumption patterns application of an efficient and powerful optimization method is a suitable solution for system planning engineering.

## 2- APPROACH TO QUANTIFY THE BENEFITS OF DG

In order to evaluate and quantify the benefits of distributed generation, suitable mathematical models must be employed along with distribution system models and power flow calculations to arrive at indices of benefits. Among the many benefits three major ones are considered: Voltage profile improvement, line loss reduction and line transmission apparent power improvement index.

### 2-1-LINE LOSS REDUCTION INDEX (LLRI)

Another major benefit offered by installation of DG is the reduction in electrical line losses [3]. By installing DG, the line currents can be reduced, thus helping reduce electrical line losses. The proposed line loss reduction index (LLRI) is defined as:

$$LLRI = \frac{LL_{W/DG}}{LL_{WO/DG}} \quad (1)$$

Where, $LL_{W/DG}$ is the total line losses in the system with the employment of DG and $LL_{WO/DG}$ is the total line losses in the system without DG and it can be:

$$LL_{W/DG} = 3\sum_{i=1}^{M} I_i^2 R_i D_i \quad (2)$$

Where, $I_i$ is the per unit line current in distribution line i with the employment of DG, $R_i$ is the line resistance (pu/km), $D_i$ is the distribution line length (km), and M is the number of lines in the system. Similarly, $LL_{WO/DG}$ is expressed as:

$$LL_{W/DG} = 3\sum_{i=1}^{M} I_i^2 R_i D_i \quad (3)$$

Advanced Computing: An International Journal ( ACIJ ), Vol.3, No.3, May 2012

Where, $I_i$ is the perunit line current in distribution line i without DG. Based on this definition, the following attributes are: LLRI < 1 DG has reduced electrical line losses, LLRI = 1 DG has no impact on system line losses, LLRI > 1 DG has caused more electrical line losses. This index can be used to identify the best location to install DG to maximize the line loss reduction. The minimum value of LLRI corresponds to the best DG location scenario in terms of line loss reduction.

### 2-2-VOLTAGE PROFILE IMPROVEMENT INDEX (VPII)

The inclusion of DG results in improved voltage profile at various buses. The Voltage Profile Improvement Index (VPII) quantifies the improvement in the voltage profile (VP) with the inclusion of DG [3]. It is expressed as:

$$VPII = \frac{VP_{W/DG}}{VP_{WO/DG}} \quad (4)$$

Based on this definition, the following attributes are: VPII < 1, DG has not beneficial, VPII = 1, DG has no impact on the system voltage profile, VPII > 1 DG has improved the voltage profile of the system. where, $VP_{W/DG}$, $VP_{WO/DG}$ are the measures of the voltage profile of the system with DG and without DG, respectively. The general expression for VP is given as:

$$VP = \sum_{i=1}^{N} V_i L_i K_i \quad (5)$$

Where, $\sum_{i=1}^{N} K_i = 1$, $V_i$ is the voltage magnitude at bus i in per-unit, $L_i$ is the load represented as complex bus power at bus i in per-unit, $K_i$ is the weighting factor for bus i, and N is the total number of buses in the distribution system. The weighting factors are chosen based on the importance and criticality of different loads.

### 2-3-Line Transmission Apparent Power Improvement Index (LTAPII)

Another advantage of DG is the transmission reduction of active and reactive powers. This increases the lines capacity, and prevents the construction and development of new lines and other facilities such as: transmission and distribution substations and therefore reduce related costs [4], It is expressed as:

$$LTAPII = \frac{LTAP_{W/DG}}{LTAP_{WO/DG}} \quad (6)$$

Based on this definition, the following attributes are: LTAPII>1, DG has not beneficial, LTAPII=1, DG has no impact on system transmission active and reactive powers, LTAPII<1, DG has improved transmission active and reactive powers. Where, $LTAP_{WO}/DG$, $LTAP_{W}/DG$ is the total line transmission apparent power without DG and with DG respectively.

$$LTAP = \sum_{i=1}^{M} I_i \cdot V_j \quad (7)$$

$V_j$ is the voltage magnitude at bus j in per units, $I_i$ is the per unit line current in distribution line i with the employment of DG and without DG.

### 3- THE OBJECTIVE FUNCTION

The proposed work aims at minimizing the combined objective function designed to reduce power loss, improve voltage profile and also increasing system load ability performance with optimum allocation of distributed generations. The main objective function is defined as (8):

$$BI = \left((BW_{VPI}).(VPII) + \left(\frac{BW_{LLR}}{LLRI}\right) + (BW_{LTAP}).(LTAPII)\right) \quad (8)$$

BI, is a complex index in order to quantify some of the benefits of DG.

$$BW_{VPI} + BW_{LLR} + BW_{LTAP} = 1 \quad (9)$$

Which $BW_{VPI}$, $BW_{LLR}$, $BW_{LTAP}$ are the weighting factors, voltage profile improvement index, line loss reduction index, and line-capacity increase index respectively. In order to apply genetic algorithm in allocation of local power stations, it is necessary to mention a few points in objective function: network losses should be reduced and network voltage buses be improved. $P_{Loss}$ is calculated using (10):

$$P_{Loss} = \sum_{i=1}^{n} P_{Gi} - \sum_{i=1}^{n} P_{Di} \quad (10)$$

Where :

n = number of network buses, $P_{Gi}$= power generation in bus i, $P_{Di}$= power consumption in bus i, $P_{Loss}$= active network losses

The problem constraints are:

1) Active power generation limitations $\quad P_{Gi}min < P_{Gi} < P_{Gi}max$
2) Reactive power generation limitations $\quad Q_{Gi}min < Q_{Gi} < Q_{Gi}max$
3) Lines passing flow through limitations $\quad P_{ij}min < P_{Gij} < P_{ij}max$
4) Voltage buses limitations $\quad V_i min < V_i < V_i max$

Objective function should be optimal, considering technical constraints.

## 4-GENETIC ALGORITM METHOD IN OPTIMAL ALLOCATION OF DISTRIBUTED GENERATION IN DISTRIBUTION SYSTEM

Genetic algorithm search method is based on natural selection and genetic mechanism. In several cases genetic algorithm is different from conventional optimization methods such as gradient methods, linear programming Genetic algorithm.

1- Works with a set of encoded parameters.
2- Starts from a parallel set of points instead of one point and the probability of reaching to the false optimum point is low.
3- Uses the original data of the objective function [5].

## 5-GA FOR PLACEMENT (GENETIC ALGORITM)

The use of GA for DG placement requires the determination of six steps as shown in Fig-5-a:

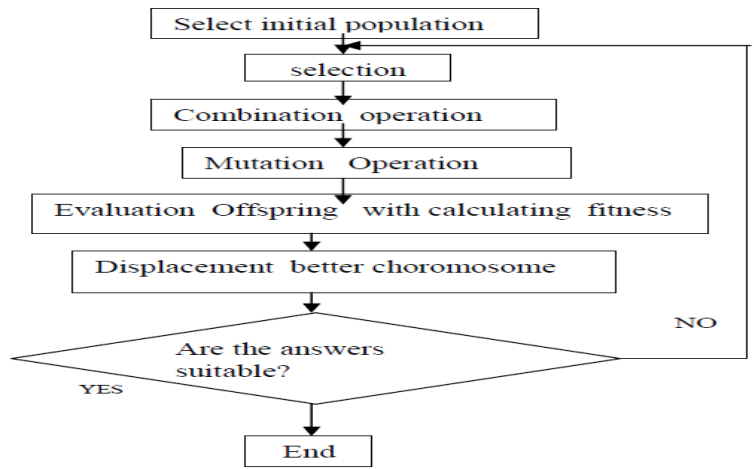

Fig-5-a

1-To install DG, the initial population are the primary proposed locations.
2-In the selection stage half of the chromosomes that have lower costs (better chromosomes) should be selected to produce their offspring.
3- In integration stage parent combination to produce offspring has been done. At this stage, for each of two selected parents we will have two children.
4- In mutation stage in the new created population that includes parents and children are mutated. At this stage designs that have a lower cost comparing to initial chromosomes, may be created.
5-In displacement stage, chromosomes with the lowest cost are selected as best chromosomes. This is the same chromosome that mutation operation is not occurred on it.
6- In continue the convergence of the answers will be checked out .The number of iterations in each step is checked in order to show that we have got the same answers or not? If we're not reaching; algorithm will return in the second stage otherwise the algorithm ends [6].

## 6-SIMULATION RESULTS ON THE IEEE 30 BUS NETWORK

The distribution system used in this paper is depicted in Fig-6-a. It is a balanced three-phase loop system that consists of 30 nodes and 32 segments. It is assumed that all the loads are fed from the substation located at node l. The loads belonging to one segment are placed at the end of each segment. The system has 30 loads totaling 4.43 MW and 2.72 Mvar, real and reactive power loads respectively.
The specifications are:
$V_{base} = 6.5kv \quad S_{base} = 10MVA$

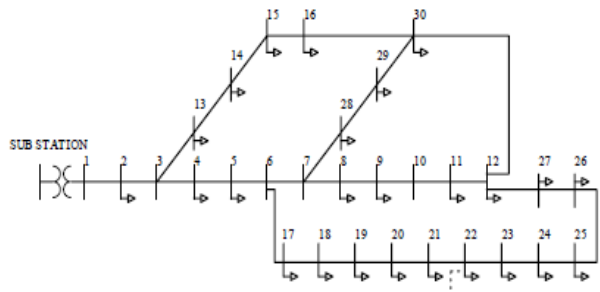

Fig-6-a: single line diagram of the test distribution system.

The whole data for this system are in reference [7] ,[8], [9].

Table-6-a:Comparison between total Active and Reactive Losses, Lines Current 1 to 3 in main feeder and Voltage regulation by using DG and without DG.

| quantity under study | Without DG | With DG | percentage reduction% |
|---|---|---|---|
| Active Losses (kw) | 380 | 71 | 81.3 |
| Reactive Losses (kvar) | 105 | 11.1 | 89.4 |
| Line current 1(A) | 488.5 | 103.5 | 78.8 |
| Line current 2(A) | 473.2 | 87.8 | 81.4 |
| Line current 3(A) | 385.5 | 8.8 | 97.7 |
| Voltage Regulation % | 9.86 | 0.758 | 92.3 |

The results by using two DGs in the buses 7 and 23 with the rating 1.75 MW and 1 MVAR is in Table-6-a.

Table-6-b: Comparison between VPII, LLRI, LTAPII indices, With employment DG and without DG.

| VP/w DG | 0.170 | LL/w DG | 0.0011 | LTAP/wo DG | 1.4892 |
|---|---|---|---|---|---|
| VP/wo DG | 0.161 | LL/wo DG | 0.0111 | LTAP/w DG | 0.2369 |
| VPII | 1.05 | LLRI | 0.0961 | LTAPII | 0.1591 |

Fig-6-c shows variation of improvement in voltage profile at all network buses with DG. The reduction in line

losses and improvement in Line Transmission Apparent Power is evident after connecting DG as shown in Table-6-b.

The current variation in main feeder, lines 1 to 3 by using DG and without DG are shown in Fig-6-b.

The voltage variation for 30 bus network by using DG and without DG are shown in Fig-6-c.

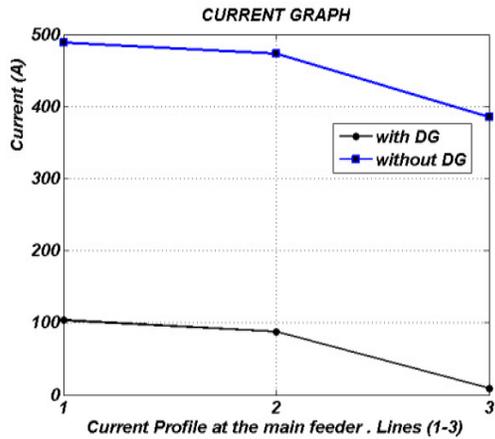 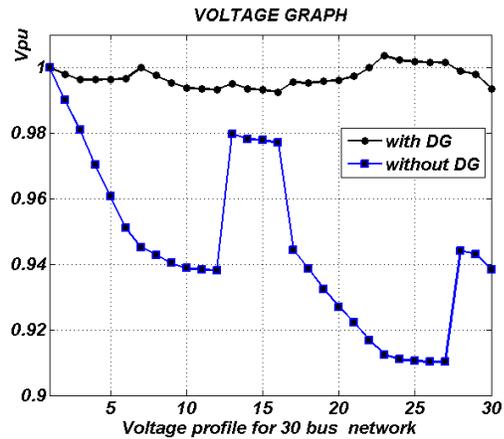

Fig-6-b                                                                 Fig-6-c

Table-6-c: Comparison between Total Active and Reactive Losses, Line Currents 1 to 3 in main feeder and Voltage regulation by using DG and without DG.

| quantity under study | Without DG | With DG | percentage reduction% |
|---|---|---|---|
| Active Losses ( kw) | 380 | 60 | 84.2 |
| Reactive Losses (kvar) | 105 | 9.2 | 91.2 |
| Line current 1(A) | 488.5 | 102.6 | 79 |
| Line current 2(A) | 473.2 | 87 | 81.6 |
| Line current 3(A) | 385.5 | 9 | 97.66 |
| Voltage Regulation % | 9.86 | 1.09 | 88.94 |

The results by using four DG at buses 6, 10, 21 and 25 with the rating 0.875 Mw and 5 Mvar in Table-6-c.

Table-6-d: Comparison between VPII, LLRI, LTAPII indices, With employment DG and without DG.

| VP/w DG | 0.173 | LL/w DG | 0.00070 | LTAP/wo DG | 1.4936 |
| --- | --- | --- | --- | --- | --- |
| VP/wo DG | 0.165 | LL/wo DG | 0.0105 | LTAP/w DG | 0.0453 |
| VPII | 1.05 | LLRI | 0.0668 | LTAPII | 32.96 |

Fig-6-e shows variation of improvement in voltage profile at all network buses with DG. The reduction in line losses and improvement in Line Transmission Apparent Power is evident after connecting DG as shown in Table-6-d.

The current variation in main feeder, lines 1 to 3 by using DG and without DG are shown in Fig-6-d.

The voltage variation for 30 bus network, by using DG and without DG are shown in Fig-6-e.

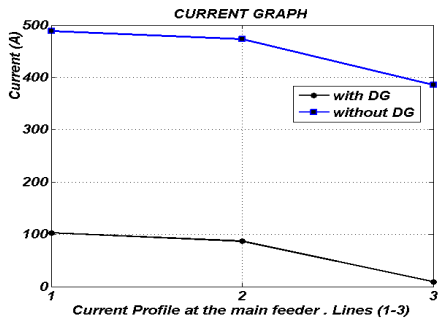
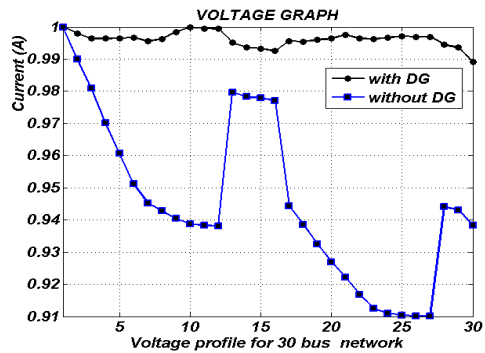

Fig-6-d　　　　　　　　　　　　　　　　Fig-6-e

## 7-SIMULATION RESULTS ON THE IEEE 34 BUS NETWORK

The distribution system used in this paper is depicted in Fig-7-a. It is assumed that all the loads are fed from the substation located at node l. The loads belonging to one segment are placed at the end of each segment. The system has 30 loads totaling 4.613 MW and 2.873 Mvar, real and reactive power loads respectively. The whole data for this system are in reference [10].

The specifications are:

$V_{base} = 6.5 kv$　　　　$S_{base} = 10 MVA$

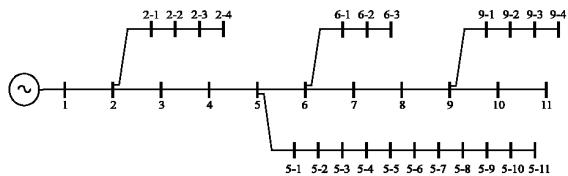

Fig-7-a .single line diagram of the test distribution system

Table-7-a: Comparison between total Active and Reactive Losses, Line Currents 1 to 3 in main feeder and Voltage regulation by using DG and without DG.

| quantity under study | Without DG | With DG | percentage reduction % |
|---|---|---|---|
| Active Losses ( kw) | 213.8 | 47.7 | 77.68 |
| Reactive Losses (kvar) | 62.7 | 11.5 | 81.65 |
| Line current 1(A) | 503.7 | 203.87 | 59.52 |
| Line current 2(A) | 479.5 | 179.8 | 62.5 |
| Line current 3(A) | 455.2 | 155.7 | 65.8 |
| Voltage Regulation % | 5.95 | 1.9827 | 66.67 |

The results by using one DGs in the bus 27 with the rating 2.75 Mw and 1.65 Mvar is in Table-7-a.

Table-7-b: Comparison between VPII, LRI, LTAPII indices, With employment DG and without DG.

| VP/w DG | 0.1800 | LL/w DG | 0.0016 | LTAP/wo DG | 1.6332 |
|---|---|---|---|---|---|
| VP/wo DG | 0.1750 | LL/wo DG | 0.0071 | LTAP/w DG | 0.5419 |
| VPII | 1.0285 | LLRI | 0.2230 | LTAPII | 0.3318 |

Figure 7 shows variation of improvement in voltage profile at all network buses with DG. The reduction in line losses and improvement in Line Transmission Apparent Power is evident after connecting DG as shown in Table-7-b.

The current variations in main feeder, lines 1 to 11 by using DG and without DG are shown in Fig-7-b.

The voltage variations for 34 bus network by using DG and without DG are shown in Fig-7-c.

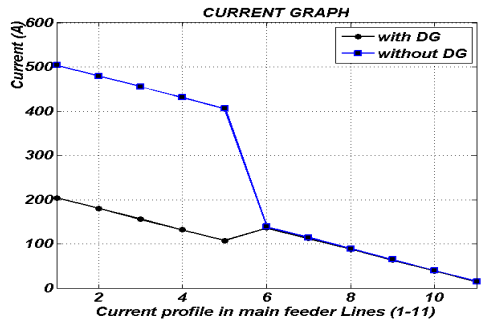 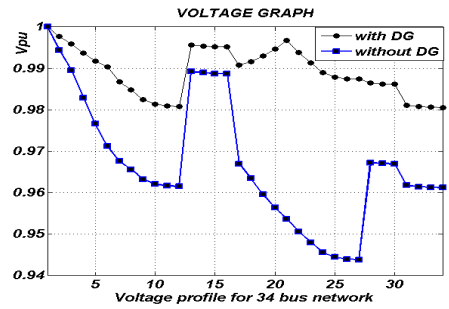

Fig-7-b                                                                 Fig-7-c

Table-7-c: Comparison between total Active and Reactive Losses, Line Currents 1 to 3 in main feeder and Voltage regulation by using DG and without DG.

| quantity under study | Without DG | With DG | percentage reduction% |
|---|---|---|---|
| Active Losses  ( kw) | 213.8 | 17.6 | 91.76 |
| Reactive Losses (kvar) | 62.7 | 3.6 | 94.25 |
| Line Current 1(A) | 503.7 | 97.21 | 80.7 |
| Line Current 2(A) | 479.5 | 75.35 | 84.28 |
| Line Current 3(A) | 455.2 | 55.34 | 87.84 |
| Voltage Regulation % | 5.95 | .3227 | 94.57 |

The results by using two DG in the buses 8, 23 with rating 2 Mw and 1 Mvar is in Table-7-c.

Table-7-d: Comparison between VPII, LLRI, LTAPII indices, With employment DG and without DG.

| VP/w DG | 0.1815 | LL/w DG | 0.00058 | LTAP/wo DG | 1.6332 |
|---|---|---|---|---|---|
| VP/wo DG | 0.1750 | LL/wo DG | 0.0071 | LTAP/w DG | 0.2322 |
| VPII | 1.0376 | LLRI | 0.0821 | LTAPII | 0.1422 |

Fig-7-d shows variation of improvement in voltage profile at all network buses with DG. The reduction in line losses and improvement in Line Transmission Apparent Power is evident after connecting DG as shown in Table-7-d.

The current variations in main feeder, lines1 to 11 by using DG and without DG are shown in Fig-7-c.

The voltage variations for 34 bus network by using DG and without DG are shown in Fig-7-d.

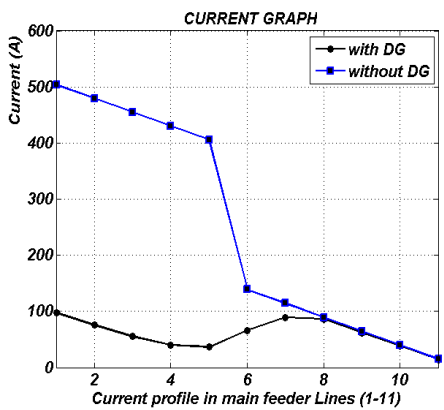
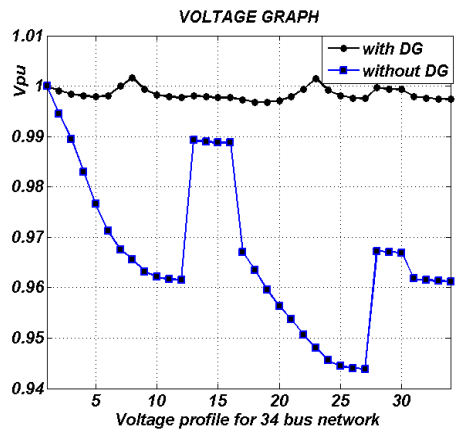

Fig-7-c                          Fig-7-d

## 8-SIMULATION RESULTS ON THE IEEE 9 BUS NETWORK

The distribution system used in this paper is depicted in Fig-8-a. The system has 9 loads totaling 12.368 MW and 4.186 Mvar, real and reactive power loads respectively. The specifications are:

$$V_{base} = 6.5 kv \quad S_{base} = 10 MVA$$

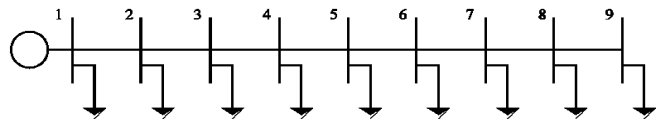

Fig-8-a .single line diagram of the test distribution system

The whole data for this system are in reference [10].

Table-8-a: Comparison between total Active and Reactive Losses, Line Currents 1 to 9 in feeder and Voltage regulation by using DG and without DG.

| quantity under study | Without DG | With DG | percentage reduction% |
|---|---|---|---|
| Active Losses ( Kw) | 438.2 | 38.2 | 91.3 |
| Reactive Losses (Mvar) | 616.8 | 75.9 | 87.7 |
| Line Current 1(A) | 1215 | 603.4 | 50.33 |
| Line Current 2(A) | 1047.7 | 436 | 58.38 |
| Line Current 3(A) | 955 | 343.7 | 64 |
| Line Current 4(A) | 791 | 181.9 | 77 |
| Line Current 5(A) | 602.6 | 80.8 | 86.6 |
| Line Current 6(A) | 446.3 | 185 | 58.55 |
| Line Current 7(A) | 371 | 245 | 34 |
| Line Current 8(A) | 261 | 237.3 | 9 |
| Line Current 9(A) | 165 | 149.7 | 9.27 |
| Voltage Regulation % | 14.65 | 3.62 | 75.3 |

The results in use of one DG at bus 7 with capacity 6 Mw and 2 Mvar is in Table-8-a.

Table-8-b: Comparison between VPII, LLRI, LTAPII indices, With employment DG and without DG.

| VP/w DG | 0.4314 | LL/w DG | 0.0048 | LTAP/wo DG | 2.1321 |
|---|---|---|---|---|---|
| VP/wo DG | 0.4160 | LL/wo DG | 0.0198 | LTAP/w DG | 0.5832 |

| VPII | 1.0370 | LLRI | 0.2410 | LTAPII | 0.2735 |

Fig-8-c shows variation of improvement in voltage profile at all network buses with DG. The reduction in line

losses and improvement in Line Transmission Apparent Power is evident after connecting DG as shown in Table-8-b.

The voltage variations for 9 bus network by using DG and without DG are shown in Fig-8-c.

The current variations feeder, lines1 to 9 by using DG and without DG are shown in Fig-8-b.

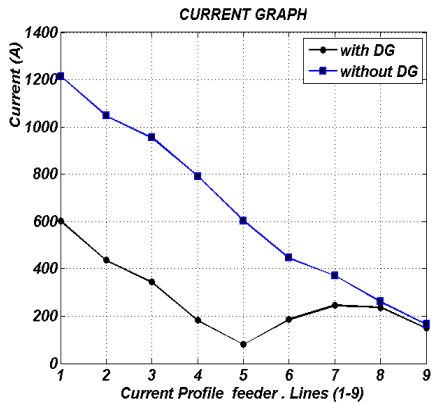
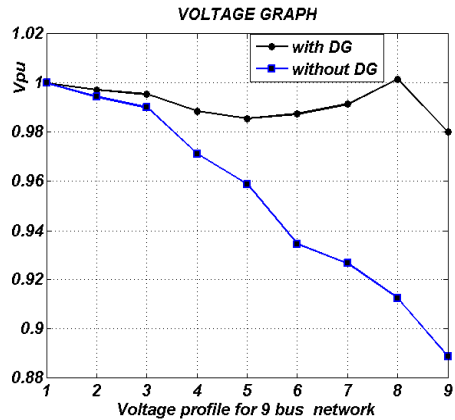

Fig-8-b                                                                    Fig-8-c

Table-8-c: Comparison between total Active and Reactive Losses, Line Currents 1 to 9 in feeder and Voltage regulation by using DG and without DG.

| quantity under study | Without DG | With DG | percentage reduction% |
|---|---|---|---|
| Active Losses ( Kw) | 438.2 | 28.6 | 93.5 |
| Reactive Losses (Mvar) | 616.8 | 58.1 | 90.58 |
| Line Current 1(A) | 1215 | 602 | 50.45 |
| Line Current 2(A) | 1047.7 | 434.6 | 58.5 |

| Line Current 3(A) | 955 | 342 | 64.2 |
| Line Current 4(A) | 791 | 180.4 | 77.2 |
| Line Current 5(A) | 602.6 | 82 | 86.4 |
| Line Current 6(A) | 446.3 | 147.8 | 66.88 |
| Line Current 7(A) | 371 | 89.5 | 75.8 |
| Line Current 8(A) | 261 | 69 | 73.5 |
| Line Current 9(A) | 165 | 141.5 | 14.2 |
| Voltage Regulation % | 14.65 | 1.56 | 89.35 |

The results in use of two DG at buses 5 and 9 with rating 3 Mw and 1 Mvar is in Table-8-c.

Table-8-d: Comparison between VPII, LLRI, LTAPII indices, With employment DG and without DG.

| VP/w DG | 0.4312 | LL/w DG | 0.0022 | LTAP/wo DG | 2.1321 |
| VP/wo DG | 0.4160 | LL/wo DG | 0.0198 | LTAP/w DG | 0.5809 |
| VPII | 1.0364 | LLRI | 0.1134 | LTAPII | 0.2725 |

Figure-8-e shows variation of improvement in voltage profile at all network buses with DG. The reduction in line losses and improvement in Line Transmission Apparent Power is evident after connecting DG as shown in Table-8-d.

The current variations feeder, lines1 to 9 by using DG and without DG are shown in Fig-8-d.

The voltage variations for 9 bus network by using DG and without DG are shown in Fig-8-e.

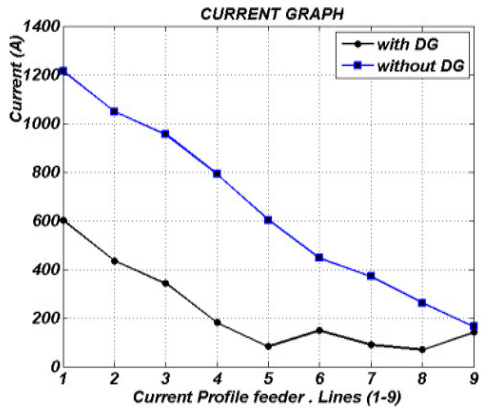 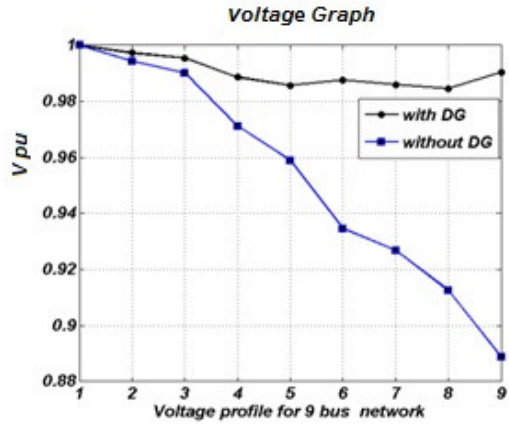

Fig-8-d            Fig-8-e

## 9-SIMULATION RESULTS ON THE IEEE 13 BUS NETWORK

The distribution system used in this paper is depicted in Fig-9-a The system has 13 loads totaling 10.536 MW and 5.962 Mvar, real and reactive power loads respectively. The whole data for this system are in refrence[11]. The specifications are:

$$V_{base} = 6.5kv \qquad S_{base} = 10MVA$$

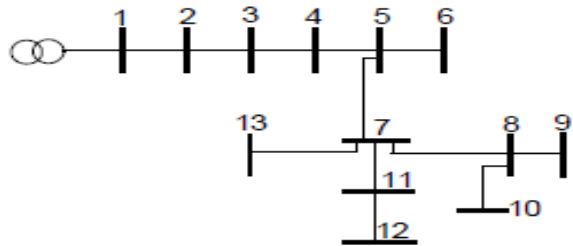

Fig-9-a. single line diagram of the test distribution system.

Table-9-a: Comparing between total Active and Reactive Losses, Line Currents 1 to 6 in main feeder and Voltage regulation by using DG and without DG.

| quantity under study | Without DG | With DG | percentage reduction% |
|---|---|---|---|
| Active Losses (kw) | 229.2 | 5.9 | 97.42 |
| Reactive Losses (kvar) | 223.5 | 21.7 | 90 |
| Line Current 1(A) | 1105 | 691 | 37.46 |
| Line Current 2(A) | 1013 | 599 | 40.86 |
| Line Current 3(A) | 940.7 | 528.44 | 43.82 |
| Line Current 4(A) | 815.6 | 406.23 | 50.19 |
| Line Current 5(A) | 55 | 53.75 | 2.27 |
| Line Current 6(A) | 691.6 | 286 | 58.64 |
| Voltage Regulation % | 7.45 | 4.32 | 42 |

The results in use of one DG at bus 7 with capacity 3.75 Mw and 2.25Mvar is in Table-9-a.

Table-9-b: Comparing between VPII, LLRI, LTAPII indices, With employment DG and without DG.

| | | | | | |
|---|---|---|---|---|---|
| VP/wDG | .3913 | LLw/DG | .0065 | LTAPwoDG | 1.967 |
| VP/woDG | .3817 | LLwo/DG | .0188 | LTAPwDG | 1.2396 |
| VPII | 1.0253 | LLRI | .3461 | LTAPII | .6302 |

Fig-9-b shows variation of improvement in voltage profile at all network buses with DG. the reduction in line losses and improvement in Line Transmission Apparent Power is evident after connecting DG as shown in Table Table-9-b.

The voltage variations for 13bus network by using DG and without DG are shown in Fig-9-b.

The current variations feeder lines1 to 12 by using DG and without DG are shown in Fig-9-c.

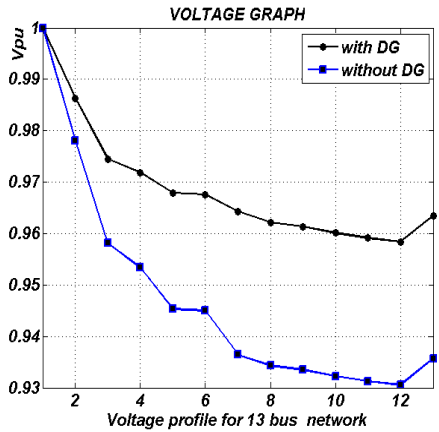
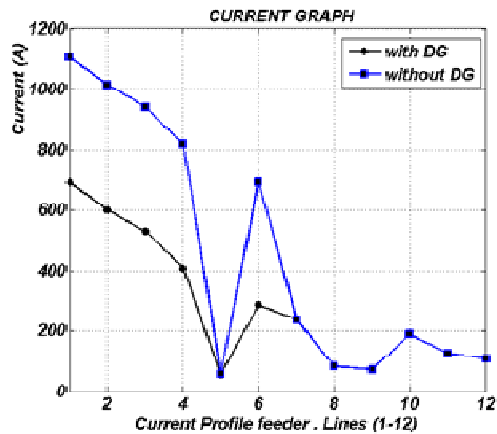

Fig-9-b　　　　　　　　　　　　　　　　　　Fig-9-c

## 10 - CONCLUSION

The introduction of DG in a distribution system offers several benefits such as relieved transmission and distribution congestion, voltage profile improvement, line loss reduction, improvement in system, and enhanced utility system reliability. This proposed work has presented an approach to quantity some of the benefits of DG namely voltage profile improvement, line loss reduction and improvement of system load ability. The results of the proposed method as applied to four IEEE network 9,13,30,34bus, clearly show that DG can improve the voltage profile and reduce electrical line losses and improve load ability index. Both ratings and locations of DG have to be considered together very carefully to capture the maximum benefits of DG. The capability of algorithm genetic is to maximise the power quality by optimizing the DG capacity.